\documentclass{aa}  
\usepackage{natbib}
\usepackage{graphicx}
\usepackage{txfonts}
\begin{document}
   \title{Active cool stars and He I 10830 \AA: the coronal connection}

   \author{J. Sanz-Forcada\inst{1}
          \and
          A.~K. Dupree\inst{2}
          }

   \offprints{J. Sanz-Forcada, \email{jsanz@laeff.inta.es}}

   \institute{Laboratorio de Astrof\'{i}sica Espacial y F\'{i}sica
     Fundamental, INTA, P.O. Box 78, E-28691 Villanueva de la
     Ca\~nada, Madrid (Spain)\\
             \email{jsanz@laeff.inta.es}
         \and
             Harvard-Smithsonian Center for Astrophysics, 60 Garden St., Cambridge, MA (USA)\\
             \email{dupree@cfa.harvard.edu}
                }

   \date{Received ; accepted }

 
  \abstract
   {The mechanism of formation of the \ion{He}{i} 10830~\AA\ triplet
     in cool stars has been subject of debate for the last 30 years. A
     relation between the X-ray luminosity and the
     \ion{He}{i} 10830~\AA\ flux was found in cool stars,
     but the dominant mechanism of formation in these stars (photoionization by coronal
     radiation followed by
     recombination and cascade, or collisional excitation in the
     chromosphere), has not yet been established.}
   {We use modern instrumentation (NOT/SOFIN) and a direct measurement
  of the EUV flux, which photoionizes \ion{He}{i}, to investigate the
     formation mechanism of the line for the
     most active stars which are frequently  excluded from
     analysis.}
   {We have observed with an unprecedented resolution ($R\sim 170,000$)
     the \ion{He}{i} 10830~\AA\ triplet in a set of 15 stars that were 
     also observed with the Extreme
     Ultraviolet Explorer (EUVE) in order to compare the line strengths with their EUV and
     X-ray fluxes.}
   {Active dwarf and subgiant stars do not exhibit a relation
  between the EUV flux and the equivalent width of the \ion{He}{i} 10830~\AA\
     line.  Giant stars however, show a positive correlation between the
  strength of the \ion{He}{i} 10830~\AA\ absorption and the EUV and
  X-ray fluxes.  The strength of the \ion{C}{iv} 1550~\AA\ emission
  does not correlate with coronal fluxes in this sample of 15 stars.}  
   {Active dwarf stars may have high chromospheric densities thus
   allowing collisional excitation to dominate
   photoionization/recombination
processes in forming the \ion{He}{i} 10830~\AA\ line. Active giant stars
possess lower gravities, and lower chromospheric 
densities than dwarfs, allowing for photoexcitation processes to
become important.  Moreover, their  extended chromospheres allow
for scattering of infrared continuum radiation, producing  strong
absorption in \ion{He}{i} and tracing wind dynamics.}

   \keywords{Stars: activity -- Stars: late-type -- Line: formation --
     Stars: chromospheres -- Infrared: stars -- X-rays: stars }

   \maketitle
%

\section{Introduction}
\begin{figure}
   \centering
   \hspace{10mm}
   \includegraphics[width=0.45\textwidth]{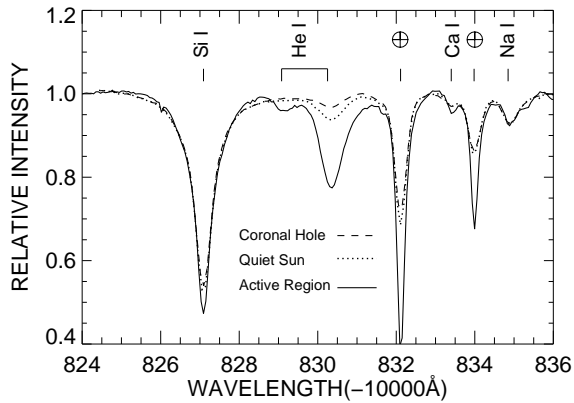}
   \caption{Spectra of different solar regions: coronal hole, quiet
     Sun, and an active region  \citep[from][with spectra from an active
     region courtesy of M. Penn]{dup96}.} 
   \label{fig:solar}
\end{figure}

\begin{figure}
   \centering
   \hspace{10mm}
   \includegraphics[width=0.45\textwidth]{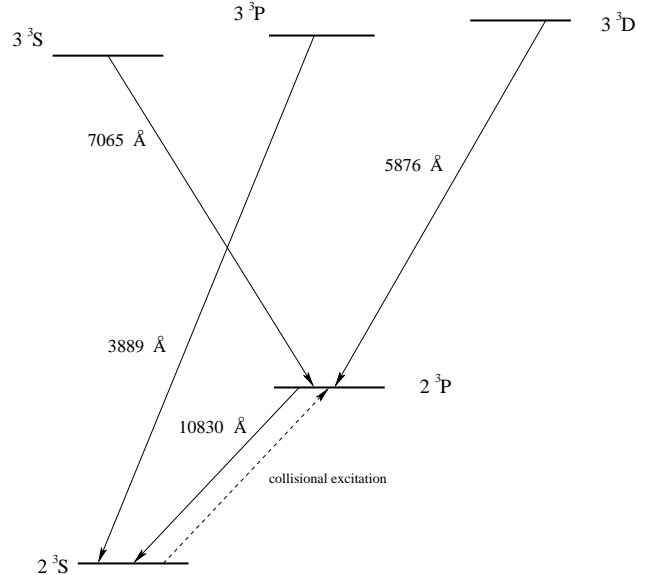}
   \caption{Energy level diagram of the triplet series in \ion{He}{i}
     showing the 10830~\AA\ transition. The broken arrow indicates 
     collisional excitation is possible in regions with high density.} 
   \label{fig:levels}
\end{figure}

Observations of the solar corona and chromosphere reveal that
regions with copious X-ray emission (emitted by  bright points or active regions in
the corona) also have enhanced \ion{He}{i} 10830~\AA\ absorption
that arises in the chromosphere
(Fig.~\ref{fig:solar}). Conversely, 
chromospheric regions located below solar coronal holes, where X-rays
are diminished, show weakened \ion{He}{i} absorption \citep{zir75,she81,dup96}.  
The \ion{He}{i} 10830~\AA\ line, actually a triplet (10829.081~\AA,
10830.250~\AA, and 10830.341~\AA), is a transition between the
lower, metastable level (2$^3$S) in the triplet series of \ion{He}{i} and the
2$^3$P level (Fig.~\ref{fig:levels}). 
The metastable 2$^3$S level can be populated only
through collisional excitation from the ground level or through
recombination and deexcitation from upper levels. 
Two mechanisms have been proposed   for the population of the lower  
2$^3$S level. In the photoionization-recombination (PR) mechanism \citep{gol39}, 
X-rays and EUV radiation ($\lambda <$504~\AA) from the corona
photoionize the neutral helium from the ground state; photoionization 
is followed by recombination, 
and the electrons cascade to populate the lower
levels of \ion{He}{i}, especially the metastable 2$^3$S level.  
Scattering of the local infrared continuum produces
an absorption line.  Models
suggest \citep{and97} the PR process
is important in the Sun at temperatures $<$10,000~K.  
However, the opacity of the chromosphere can limit
the efficiency of the photoionization 
mechanism. 
In an alternative mechanism, electron collisions from the ground
and the  metastable 2$^3$S  level of 
\ion{He}{i} dominate photoionization. In this case, a
temperature $\sim$ 20,000-30,000~K is required, and 
high densities in the chromosphere and transition region enhance the
process.
A combination of the two processes can exist as well. 
Arguments for each of the mechanisms can be found in
\citet{zir82,sim82,smi83,wol84,obr86, zar86,lan95,and95,and97,piet04}, and references therein. 

If the PR mechanism dominates the formation of the 10830~\AA\ transition, a 
correlation is expected between the strength of the helium
line and the radiation field at $\lambda <$504~\AA\ located in the EUV and X-ray
bands. Several studies  have compared the equivalent widths
(EW) of the \ion{He}{i}~10830~\AA\ line and the X-ray flux in late-type
stars in order to establish a relation between those
parameters. Theses studies show that stronger X-ray emission yields  stronger 10830~\AA\
absorption in both dwarfs (of spectral type F7 or later) and giants 
\citep{zir82,obr86,zar86}. However the
RS~CVn active binary systems 
were generally not included in these analyses.
Many of the  previous observations made use of photographic plates to 
measure the 10830~\AA\ line, and correlated the line strengths with 
the X-ray fluxes observed by 
the  Einstein satellite (IPC and HRI instruments) 
which spanned the energy range  0.1~--~4 keV. In this paper 
we present measurements of the 10830~\AA\ line in very active cool
stars and binaries taken with modern
instrumentation and high resolution.  The use of high spectral 
resolution is essential
in order to discern blends with telluric lines. Additionally,
EUV fluxes are obtained as measured directly with
the Extreme 
Ultraviolet Explorer (EUVE). Since the photoionization edge
of \ion{He}{i} is located at 504~\AA\ (0.02 keV), the EUV fluxes are 
expected to relate  closely  to the photoionization rate
of helium. 

The chromospheric line of \ion{He}{i} at 10830~\AA\ can indicate bulk mass motions
in the atmospheres of cool luminous stars.  The lower level of the
transition is metastable, and a large population can build
up in this level. If the gas itself is moving outward in a wind, the 
helium in the metastable level  
in the outflow scatters  photospheric infrared radiation,
and the line profile can reveal the dynamics of the atmosphere. If there
is a significant contribution by the photoionization-recombination
process, then the level population is independent of the
local thermodynamic conditions in the chromosphere.  In all
cases, detailed modeling demonstrates \citep[see][]{dup92} that
in luminous stars, the Helium atom is formed further out in 
the atmosphere than other optical diagnostics of mass flow 
such as H$\alpha$ and \ion{Ca}{ii} H \& K lines, making it an 
extremely sensitive probe of the acceleration region 
of a stellar wind. Even in a dwarf star, such
as the Sun, models demonstrate that the 10830~\AA\ line
is formed above the \ion{Ca}{ii} K core and H$\alpha$ \citep{avr98}
so that this Helium transition is more favorable for detecting
regions where acceleration is likely to occur.   Most importantly, observations
of the 10830~\AA\ line in both Sun and stars  
demonstrate that high outflow 
velocities are  observed in the line profiles of cool stars \citep[see
e.g.][]{obr86, dup92, dup96, edw03, dup05}. Given the
high spectral resolution 
of our observations, the presence of winds in the stars
of the sample can be explored. The cool giants are
especially good targets.

In Sect. 2 we  describe the observations. Sect. 3
compares the \ion{He}{i} line strengths to other parameters. Results are 
discussed in Sect. 4; conclusions can be found in 
Sect. 5.  

\section{Observations}

\begin{figure*}
   \centering
   \hspace{10mm}
   \includegraphics[width=0.9\textwidth]{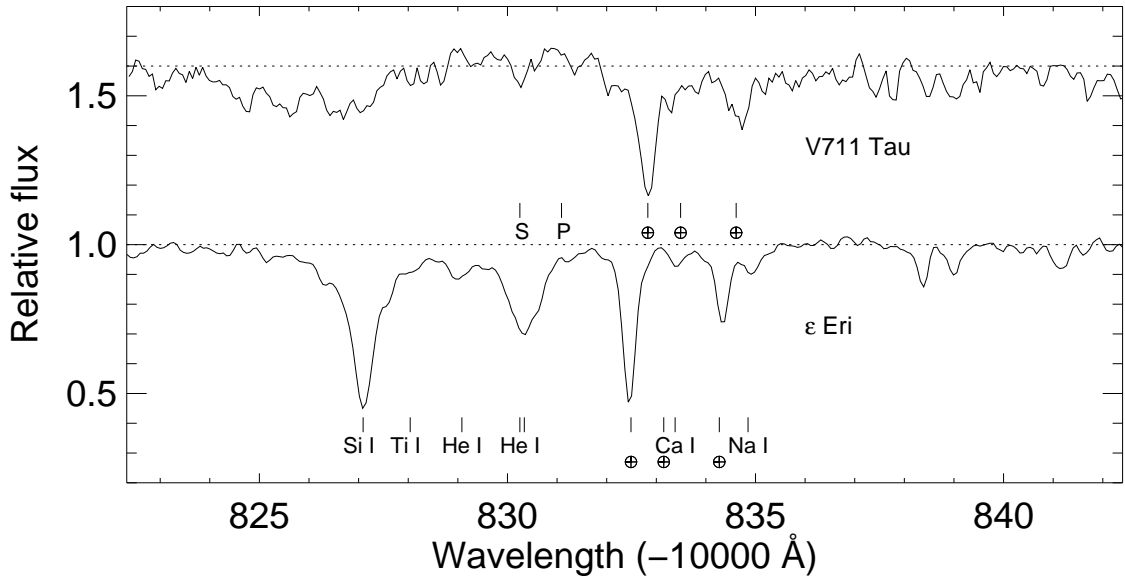}
   \caption{Spectra of $\epsilon$ Eri and V711 Tau. Positions of the
     \ion{He}{i}~10830.25 \AA\  line corresponding to
     the primary (P) and secondary (S) components of V711 Tau are
     marked. In this observation of V711 Tau (16/08/2000, 6:07 U.T.)
     the \ion{He}{i}~10830 line is slightly in emission, which
     can arise  in an extended atmosphere.} 
   \label{fig:epseri}
\end{figure*}

\begin{figure*}
   \centering
   \hspace{10mm}
   \includegraphics[width=0.9\textwidth]{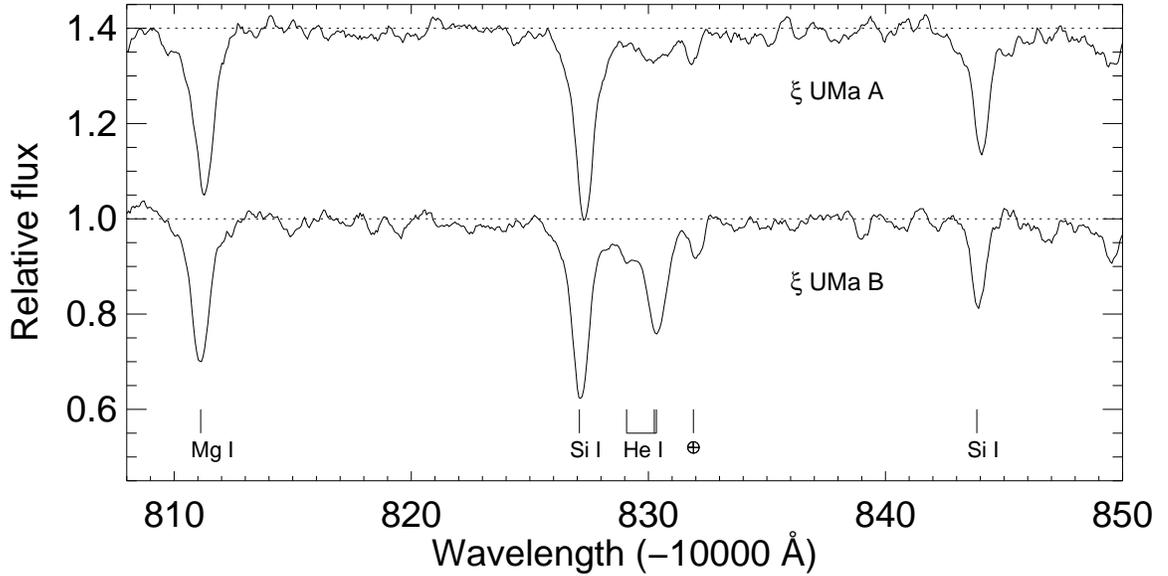}
   \caption{Spectra of $\xi$ UMa A and B. The
     \ion{He}{i}~10830\AA\  line reveals a substantial difference in activity between the
     A and B components (G0+G5 dwarf stars respectively). 
     The source of X-rays
     is the B member of this visual binary system which is itself a
     short-period (3.98 days) binary, observed pole-on.  The
     B component exhibits  the stronger \ion{He}{i} absorption.}  
   \label{fig:xiuma}
\end{figure*}

\begin{figure*}
   \centering
   \hspace{10mm}
   \includegraphics[width=0.8\textwidth]{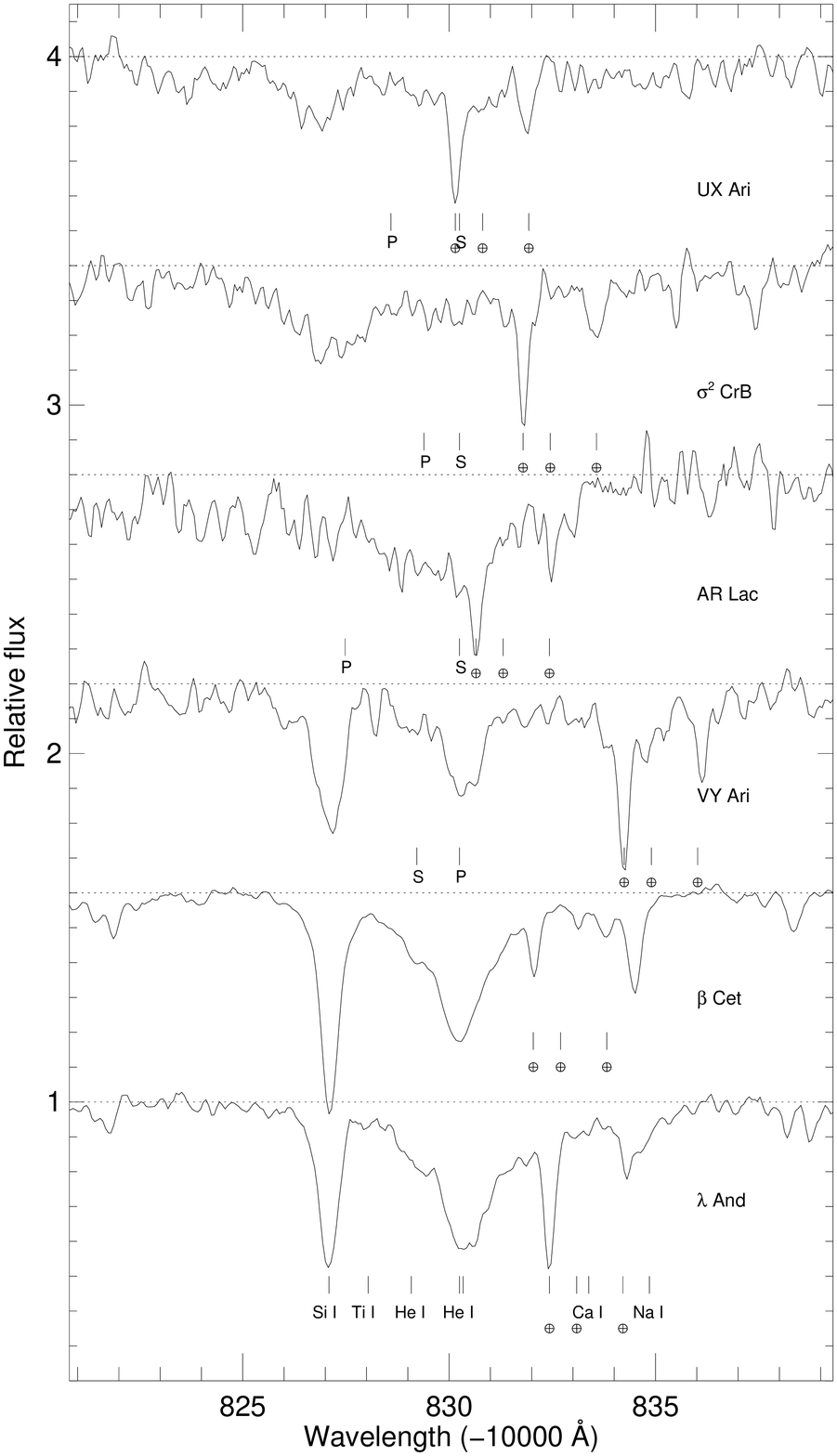}
   \caption{Spectra of several sources normalized to 1 (indicated by
     the broken lines) and offset in this plot. Positions of the
     strongest line in the \ion{He}{i} multiplet, $\lambda$10830.25
     corresponding to 
     the primary (P) and secondary (S) components of binaries are
   marked. The profile of Si I reveals 
   broadening  due to turbulence in the stellar
   photosphere and rotation. } 
   \label{fig:sevstars}
\end{figure*}

\begin{figure*}
   \centering
   \hspace{10mm}
   \includegraphics[width=0.8\textwidth]{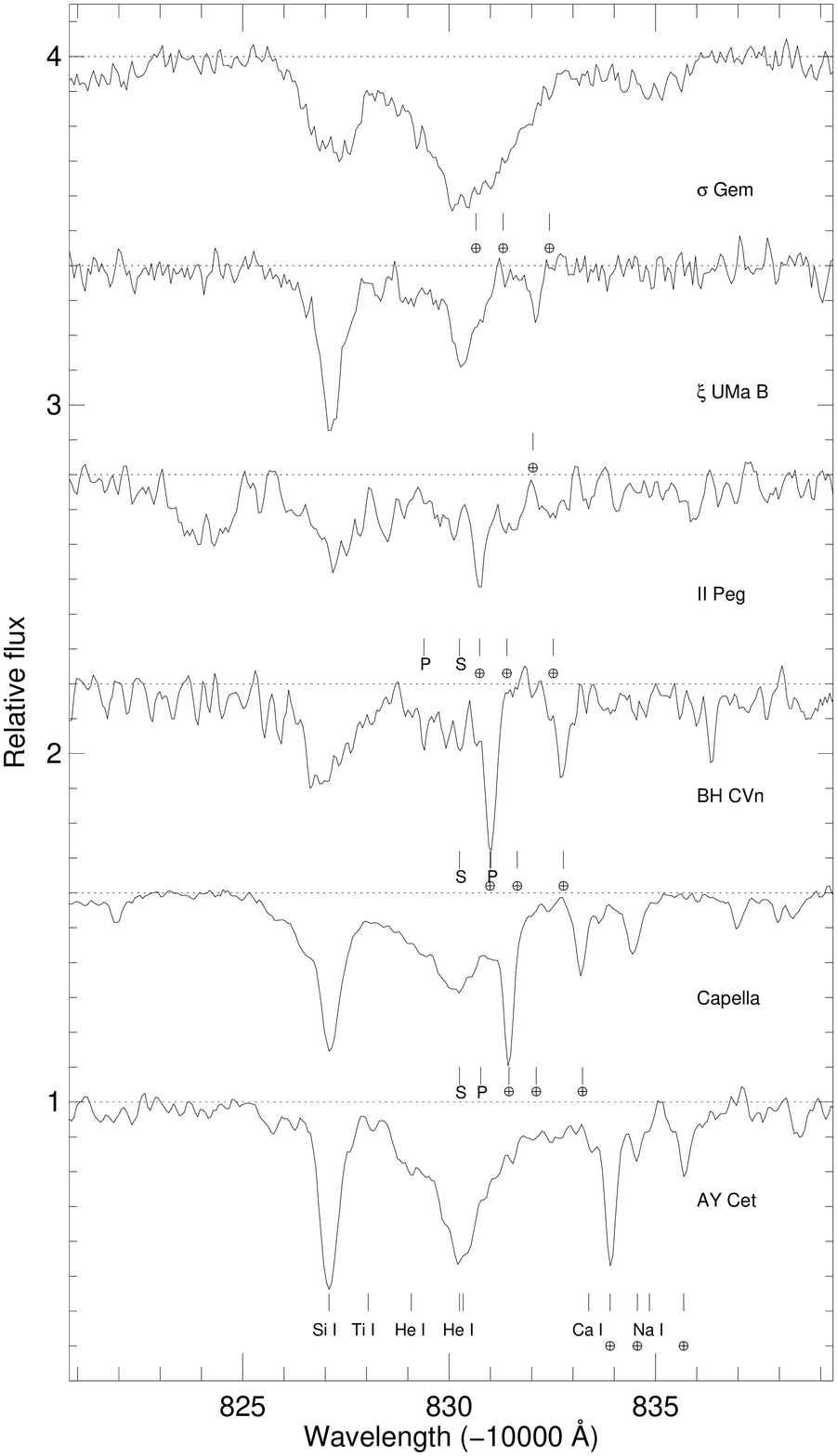}
   \caption{Same as  Fig.~\ref{fig:sevstars}.} 
   \label{fig:sevstars2}
\end{figure*}

High-resolution infrared spectra of active binary systems and
single stars (Table~\ref{fluxtable}) 
were obtained during  11--19 August 2000 at
the 2.56\,m Nordic Optical Telescope (NOT)
located at the Observatorio del Roque de Los Muchachos (La Palma,
Spain). The Soviet Finnish High Resolution Echelle
Spectrograph (SOFIN) was used to record spectra in the spectral range $\lambda
\lambda\sim$10800--10870, with a resolution of 
R$\sim$170,000. This resolution allows the \ion{He}{i} triplet
($\lambda$10829.081, $\lambda$10830.250, $\lambda$10830.341) to
be resolved and the profiles measured. It also provides clear separation from 
nearby lines, such as \ion{Si}{i} 10827.09~\AA, \ion{Ti}{i}
10828.04~\AA\ and the water vapor line at $\sim$10832.1~\AA\
(Fig.~\ref{fig:epseri}--\ref{fig:sevstars2}). Exposure times varied
from 120~s to 80~min; individual exposures were no longer than
20~min at most and depend on  stellar magnitude.
A ThAr lamp was used for the wavelength calibration. 
We have reduced the data using standard procedures within the Image
Reduction and Analysis Facility (IRAF) software
package, developed by the U.S. National Optical Astronomy Observatories
(NOAO).  Equivalent widths were measured from the normalized spectra
by fitting multiple gaussians to the blended profiles which at times
included telluric lines.
The contribution of telluric water vapor lines was subtracted
from the Helium equivalent width. 
Since spectra were obtained at different orbital  phases of the
binaries, we have
measured the equivalent widths displayed in Table~\ref{fluxtable} in
the spectra with the widest velocity separation of the
components. This ensures the least blending of  individual features.
Typical errors in the measurement of the equivalent width are less
than 15\%, and the line centers correspond to their
theoretical values within 5~km\,s$^{-1}$. We tested the line ratio
between $\lambda$10830.250 + $\lambda$10830.341 and $\lambda$10829.081
in $\epsilon$~Eri. A line ratio of 6.5$\pm$3.6 is measured, consistent
with the expected value of 8 based on {\it gf} values. $\xi$ UMa B has 
a ratio of 9.6$\pm$0.5, which is also reasonable.

Spectroscopic observations taken with the Extreme Ultraviolet Explorer (EUVE)
satellite were obtained from the MAST Archive at Space Telescope
Science Institute (STScI), and also directly from the EUVE project for
our Guest Investigator targets.  Fluxes in the
80--170~\AA\ range were calculated from the EUVE 
spectra by first summing all available spectra.  
The summed fluxes were corrected for absorption by  the interstellar medium
as explained in \citet{sanz03}. Although it would be
possible with EUVE to measure the flux close to the helium edge at
504~\AA, in 
most cases the spectra were  absorbed by the interstellar
medium  at those wavelengths  and could not give 
reliable  measurements. (The flux in the range 170--370~\AA\ can be as
much as 14 times the flux in photon units in the 80--170~\AA\ range). 
The measurement of the \ion{He}{ii}~304~\AA\ line is also possible
in most cases, but the extraction is difficult because the large
aperture is filled with airglow around this line. We treat it in 
similar fashion as sky subtraction in a slit used for optical spectroscopy.
In the case of AY Cet, where no measurement of the 304~\AA\ line was
available, we used the relation observed between values of $L_{\rm EUV}$ and
$L_{\rm He II~304}$ in the sample to calculate the expected value of $L_{\rm He
  II~304}$  (Table~\ref{fluxtable}). 
Finally, X-ray fluxes (0.1--2.4 keV) obtained with the Roentgen
Satellite (ROSAT) were taken from the Rosat All Sky Survey (RASS) data
\citep[][and references therein]{vog99}.

\begin{figure}[t]
   \centering
  \includegraphics[width=0.5\textwidth]{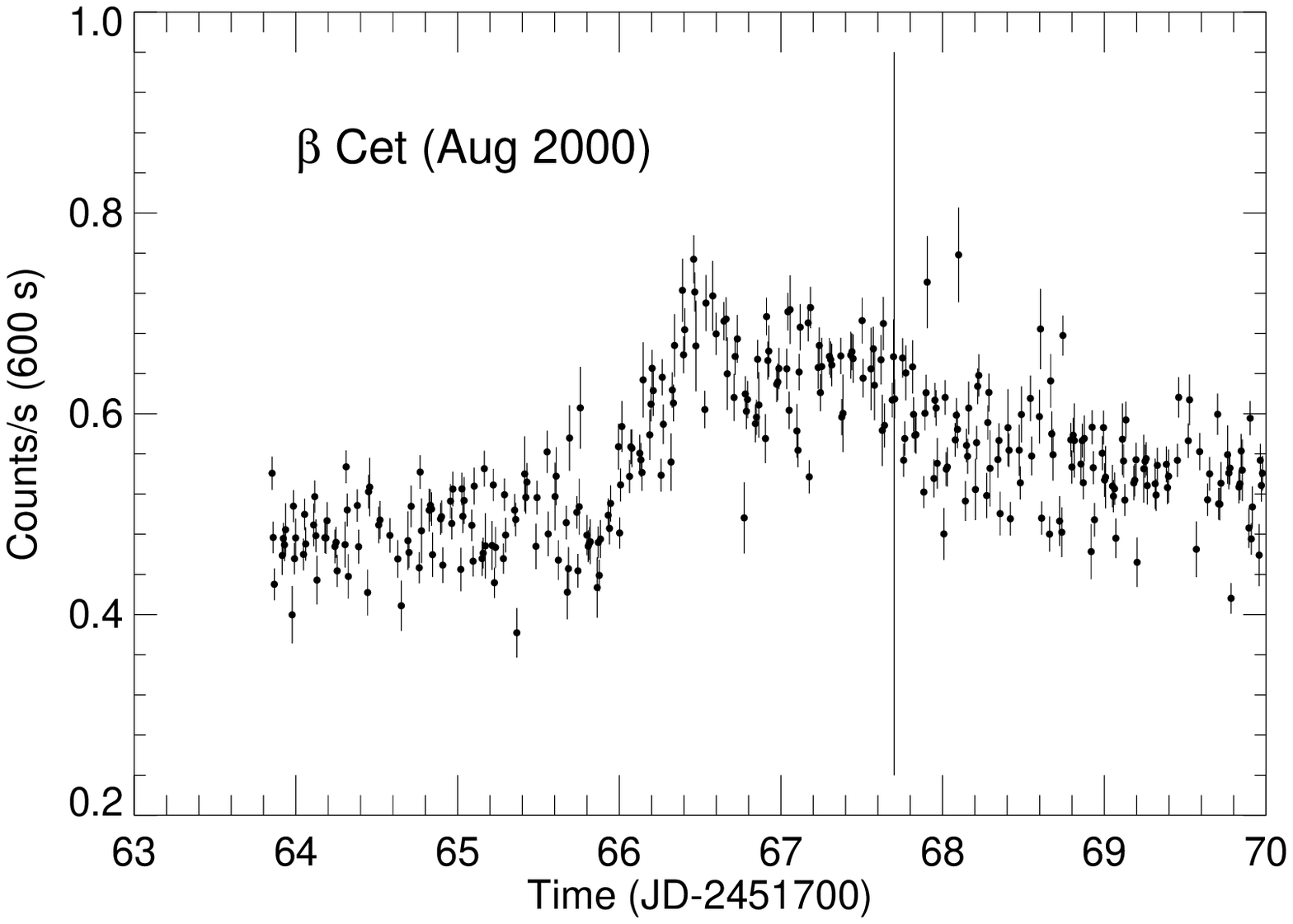}
  \includegraphics[width=0.5\textwidth]{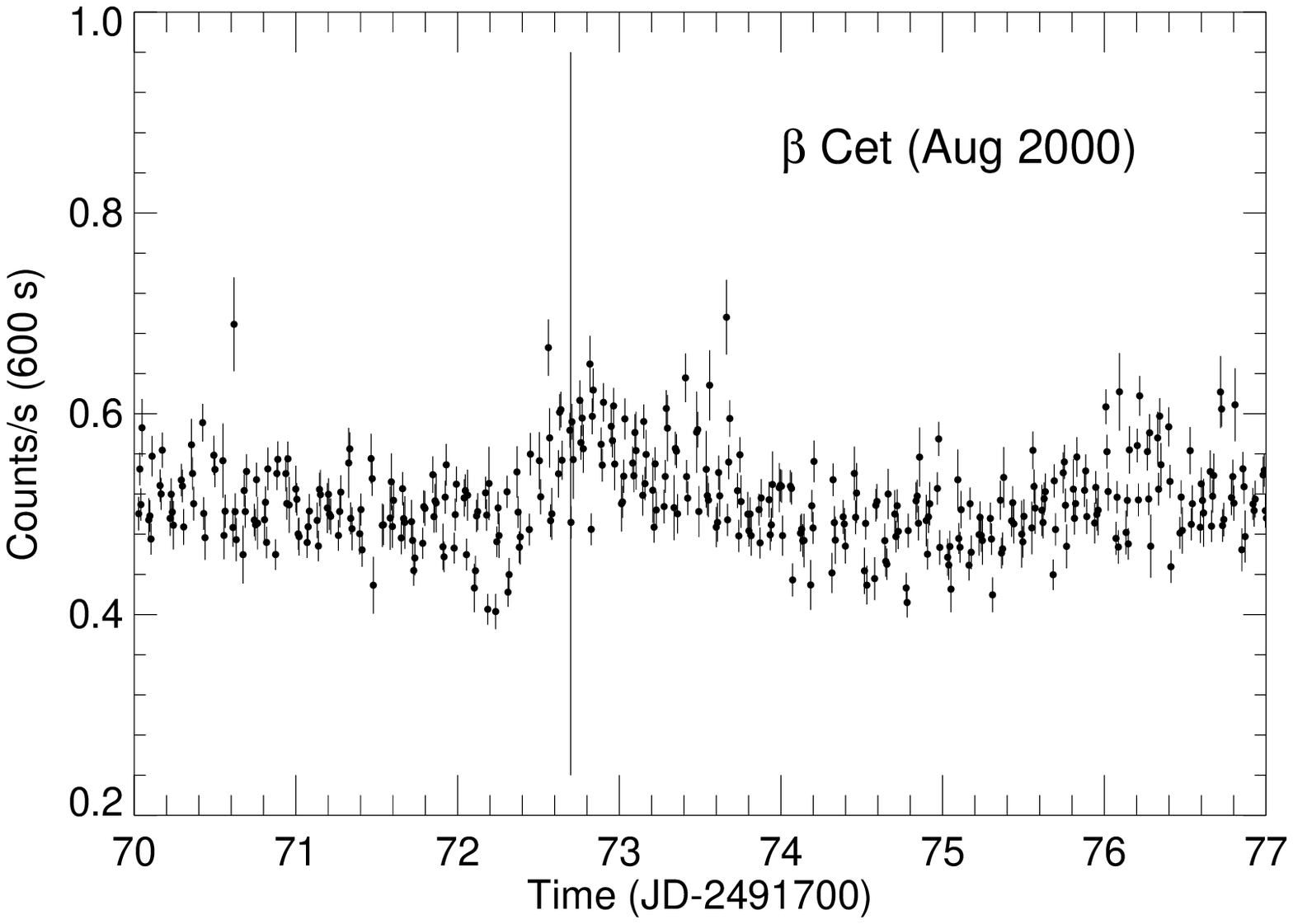}
   \caption{EUVE light curve flux variation in $\beta$ Cet over 14
     days of continuous observation. Solid lines mark the times of the
     two \ion{He}{i}
     10830~\AA\ observations reported here. The first of these, with 
     EW=0.905~\AA, corresponds to the first small EUV flare \citep[see
     also][]{sanz03}, and the second (EW=0.873) is located precisely at
     the maximum of a small EUV flare.}
    \label{fig:betcet1}
\end{figure}

Many of these targets exhibit flux variations in the EUV spectral
range. Flaring can cause 
short-lived increases that may amount to a factor of 6; some
targets  also show changes when
active regions presumably rotate into view \citep{sanz02,ost99}.
Long pointings with the EUVE satellite towards $\sigma$ Gem, V711
Tau, UX Ari \citep{sanz02} and $\beta$ Cet \citep{ayr01} 
reveal that the EUVE fluxes of these targets are modulated
by a factor of 1.2 ($\beta$ Cet), 
2 ($\sigma$ Gem), 3 (V711 Tau), and
5 (UX Ari), outside of flaring episodes (see also Fig.~\ref{fig:betcet1}).

\section{Results}

The secondary stars in binary systems
are frequently more active than the primary  due to the presence of a deeper convective
region. In many cases, our spectra show a stronger 10830~\AA\
absorption line from the secondary star. 
Several stellar parameters are compared to investigate the dependence
of the \ion{He}{i} strength on the radiation field. 
Fig.~\ref{fig:heeuv} shows the relation between 
the EUV luminosity at the star (corrected for interstellar 
absorption) and the equivalent 
width of the 10830\,\AA\ absorption line. The active dwarf stars exhibit
EUV luminosities that span a factor of about 1000, yet
the \ion{He}{i} equivalent width changes at most by a factor of 4.
The dwarfs and subgiants have absorption equivalent widths not in excess of
~400~m\AA.  An  exception is  AR~Lac, which has 2 active stars.
The measured EW lies below 800~m\AA, and its value does not appear
to be varying systematically  with the EUV flux. Very active stars such as
V711~Tau or UX~Ari have a weak 10830~\AA\ line that possibly  
contains emission in one observation of V711~Tau (Fig.~\ref{fig:epseri}). 

\begin{table*}[t]
\caption{Data for Target Stars \citep[targets selected from][]{sanz03}.\label{fluxtable}} 
\begin{center}
\setlength\tabcolsep{2.8pt}
\begin{tabular}{lrccccccccc}
\hline\hline\noalign{\smallskip}
{Star} & HD & {Spectral Type} & d & {\ion{He}{i} 10830} & {Date, U.T.}
& {log $L_{\rm EUV}$} & $F_{\rm EUV}$\footnote{Flux at Earth, corrected for interstellar absorption. EUV flux in range 80--170 \AA, X-ray flux in range 0.1--2.4 keV} & 
$F_{\rm He\,II~304}$ & {log $L_{\rm X}/L_{\rm bol}$} & {$F_{\rm C\,IV}$}\\
& & & [pc] & W$_{\lambda}$ [\AA] & [midpoint] & [erg s$^{-1}$] & [erg
  cm$^{-2}$s$^{-1}$] & [erg cm$^{-2}$s$^{-1}$] & & [erg cm$^{-2}$s$^{-1}$] \\ 
\noalign{\smallskip}
\hline
\noalign{\smallskip}
$\beta$ Cet   & 4128&K0III      & 29.4 & 0.890 & 16/08/2000, 03:48 &  29.6 & 3.56\,E--12 & 6.06\,E--13 & --5.3 & 9.64\,E--13 \\
AY Cet         & 7672&WD/G5III  & 78.5 & 0.880 & 15/08/2000, 04:05 &  30.3 & 2.67\,E--12 & (8.61\,E--13)\footnote{Calculated from the $L_{\rm EUV}$ vs. $L_{\rm He II~304}$ relation in the rest of sample} & --4.1 & 9.06\,E--13 \\
VY Ari         & 17433&K3-4V-IV + ?& 44.0 & 0.400& 14/08/2000, 04:08 &  30.2 & 7.16\,E--12& 9.65\,E--13 & --3.2& 4.70\,E--13 \\
UX Ari         &21242 &G5V/K0IV   & 50.2 & 0.240 & 16/08/2000, 04:40 &  30.5 & 10.7\,E--12& 17.9\,E--13 & --3.3& 11.5\,E--13 \\
$\epsilon$ Eri & 22049&K2V        & 3.22 & 0.310 & 13/08/2000, 05:46 &  27.5 & 2.57\,E--12& 21.4\,E--13 & --4.8 & 11.7\,E--13 \\
V711 Tau       &22468 &G5IV/K1IV  & 29.0 & 0.090 & 17/08/2000, 05:47 &  30.1 & 12.0\,E--12& 46.1\,E--13 & --3.0 & 35.6\,E--13 \\
Capella        & 34029&G1III/G8III & 12.9 & 0.630& 15/08/2000, 06:47 &  29.4 & 11.5\,E--12& 125\,E--13 & --5.3 & 416\,E--13 \\
$\sigma$ Gem   &62044 &K1III + ?  & 37.5 & 1.110 & 14/04/2000, 20:19 &  30.2 & 9.33\,E--12& 20.4\,E--13 & --4.1 & 30.7\,E--13 \\
$\xi$~UMa B   & 98230&G5V/[KV]    & 8.35 & 0.380 & 17/04/2000, 21:45 &  28.5 & 4.11\,E--12& 18.5\,E--13 & --4.3 & 10.4\,E--13 \\
BH CVn         & 118216&F2IV/K2IV  & 44.5 & 0.190 & 13/08/2000, 20:20 &  29.7 & 2.34\,E--12& 33.9\,E--13 & --4.2 & 14.5\,E--13 \\
$\sigma^2$ CrB & 146361&F6V/G0V    & 21.7 & 0.320 & 18/08/2000, 22:23 &  29.8 & 12.0\,E--12& 63.8\,E--13 & --3.4 & 26.1\,E--13\\
AR Lac         & 210334&G2IV/K0IV  & 42.0 & 0.760 & 14/08/2000, 02:20 &  30.0 & 5.21\,E--12& 14.5\,E--13 & --3.4 & 14.9\,E--13 \\
$\lambda$ And  &222107 &G8IV-III + ?& 25.8 & 1.010& 15/08/2000, 03:10 &  29.9 & 10.3\,E--12& 24.5\,E--13 & --4.3 & 30.4\,E--13 \\
II Peg         &224085 &K2IV/M0-3V & 42.3 & 0.300 & 12/08/2000, 03:45 &  30.0 & 5.12\,E--12& 11.4\,E--13 & --2.8 & 4.82\,E--13 \\
\hline
\end{tabular}
\end{center}
{$^1$ Flux at Earth, corrected for interstellar absorption. EUV flux in range 80--170 \AA, X-ray flux in range 0.1--2.4 keV \\}
{$^2$ Calculated from the $L_{\rm EUV}$ vs. $L_{\rm He\,II~304}$ relation in the rest of sample}
\end{table*}

The \ion{He}{i} $\lambda$10830  line shows a different behavior in the
5 active giants (Figs.~\ref{fig:heeuv}, \ref{fig:helx}). Here  
stronger \ion{He}{i} absorption appears 
directly correlated with  stronger EUVE or X-ray flux (correlation factor $r=0.71$).
The long-period binary Capella ($\alpha$~Aur) deserves special mention.
\citet{kat98} monitored this system over 9 years
and detected a variation of the equivalent width of the 10830~\AA\ line
with orbital phase.  At the time of our observations, the orbital
phase was 0.79 which, in conjunction with the equivalent width,
suggests
that Capella was in a moderately active phase \citep[see Fig.~4 of
][]{kat98}. The total variation in equivalent width of $\lambda$~10830
appears to
be a factor of 1.5, and its value could be as high as
0.8~\AA. Results
from the EUVE satellite show that the EUV radiation from highly
ionized \ion{Fe}{xx} through \ion{Fe}{xxiii} in Capella is modulated
also on an orbital time scale \citep{dup95}, by a factor $\sim$2,
although
the effect on the total EUV flux is mitigated by the relatively
constant flux of the 
strong \ion{Fe}{xviii} lines in the EUV region. Even when these variations
are considered, Capella remains at the low end of the
$W_\lambda$--$L_{\rm EUV}$ relation for giants shown in Fig.~\ref{fig:heeuv}.  

Broad wings in the  giants $\lambda$~And,
$\beta$~Cet, $\sigma$~Gem, and Capella indicate 
mass motions or turbulence in the upper chromosphere of the star. 
Of particular note, the He absorption extends to shorter wavelengths
to and, in some cases,  beyond the \ion{Si}{i} absorption line which lies 90 $\rm {km\
s}^{-1}$ distant.  The He line absorption profile traces out the
behavior of velocity 
in an  expanding
atmosphere.  Since giant stars have extended 
chromospheres, these velocities become comparable  to the stellar
escape velocity, and can be said to mark the presence of a stellar wind. 
At the time when
we observed the \ion{He}{i}~10830\AA\ line in $\lambda$ And, \citet{mir03} 
monitored the star 
searching for periodicity in  visible light 
and in  photospheric spot coverage. During the period of our
NOT observations of $\lambda$~And, no important 
changes occurred in the observations of \citet{mir03}, and the
\ion{He}{i}~$\lambda$10830 did not show substantial changes in
strength either.  
 
EUVE was also observing $\beta$~Cet during our NOT campaign. Between
the two NOT observations (JD 2451767.7 and 2451772.7) there were no
changes in the line profiles or intensity. The NOT spectra are simultaneous
with two local enhancements of similar strength in the EUVE light curve
\citep[Fig. 7 and][]{sanz03} that are likely due to a small flare or
to an active 
region. However, our value of $W_\lambda =$~0.890~\AA\ is larger than
that (0.3 to 0.5~\AA) measured
by \citet{obr86} for $\beta$~Cet, and the line depth also increased
substantially in our observations. These  may signal  enhanced
activity during the flaring episodes.
The star $\beta$ Cet, a presumably single  giant (K0 III) follows
the relation for giant stars that are members of RS~CVn systems with
respect to its EUV flux and $\lambda$10830 equivalent width.  Ions of
high excitation, \ion{Fe}{xviii} through \ion{Fe}{xxiii} have been
observed directly in its EUV and 
far-UV spectrum \citep{sanz03,red03,dup05b} and flaring episodes
occur \citep{ayr01}.  The star, $\beta$~Cet,  is a slow
rotator, with $v~\sin~i~=~4$~km\ s$^{-1}$ \citep{fek97}, and it is
puzzling how its 
activity appears comparable to the RS~CVn objects.  It is not
out of the question that $\beta$ Cet is observed pole-on, and in
actuality may be an active binary system.

\section{Discussion}
In this sample of active stars, 
the dwarfs and subgiants show  different behavior from  that of the
giants. In the dwarf stars, the 
increasing levels of activity (seen through the X-ray or
EUV luminosities) do not cause more substantial \ion{He}{i} 10830~\AA\
absorption. On the contrary, the absorption looks rather shallow (if
not partly in emission) for very active stars such as
V711~Tau. Modelling efforts carried out by \citet{and95} conclude 
that there is a maximum limit in the equivalent width of this line of around $\sim
400$~m\AA\ in dwarf stars. For higher levels of activity the flux in 
the line is larger 
than in the continuum, and it becomes an emission line,
resulting in a profile that is a balance of absorption and emission. 
The value of 400~m\AA\  corresponds to the maximum equivalent width
found in our sample of dwarfs and subgiants corroborating the Andretta
\& Giampapa results.  \citet{zar86} concluded
that for F, G, and K dwarfs (spectral type equal to F7 or later) 
with weak coronas ($L_X/L_{bol} < 10^{-4}$), a relation
exists between the X-ray flux and the equivalent width of the
$\lambda$10830 line.  Our sample focuses on strong X-ray emitters, and
for these objects the relation does not hold.  It would appear that
the densities in these active dwarfs become sufficiently high to
allow collisional processes to become significant.

\begin{figure}[t]
   \centering
   \hspace{10mm}
   \includegraphics[width=0.5\textwidth]{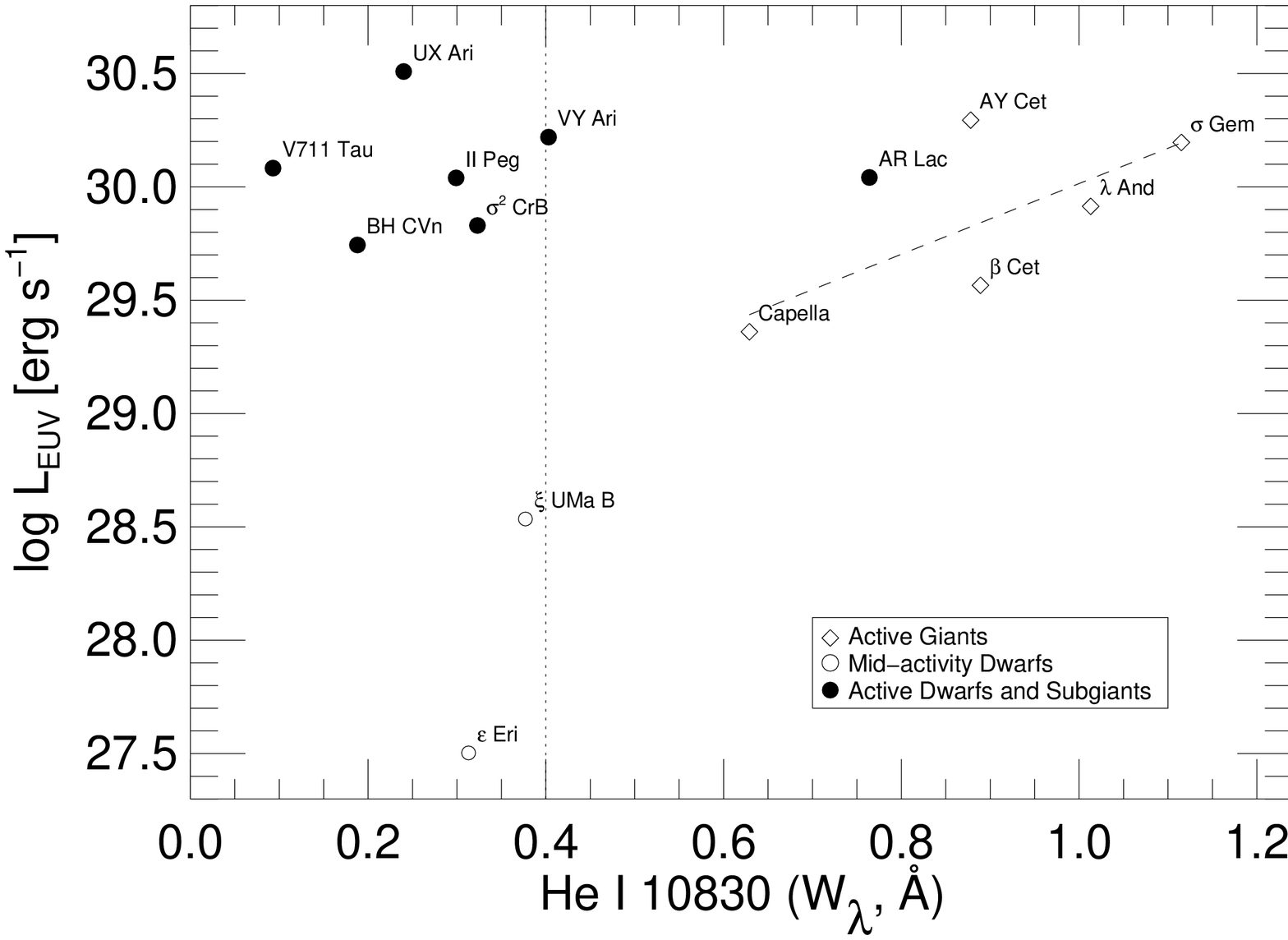}
   \caption{EUV luminosity in the range 80--170~\AA\ \citep[EUV
     luminosities from][]{sanz03} vs. the equivalent width of 
     the \ion{He}{i} 10830~\AA\ triplet. Equivalent 
     widths have typical errors of $\leq$0.1~\AA. A dashed line
     indicates the linear fit to the giants (correlation factor $r=0.71$).
     The dotted line at EW=0.4 \AA\ marks the limit suggested by
     models \citep{and95} of a single dwarf star and corresponds to
     our measured limit too. Note that AR Lac has two active stars and
     EW$\lesssim$0.8 \AA. }
   \label{fig:heeuv}
\end{figure}

\begin{figure}[t]
   \centering
   \hspace{10mm}
   \includegraphics[width=0.5\textwidth]{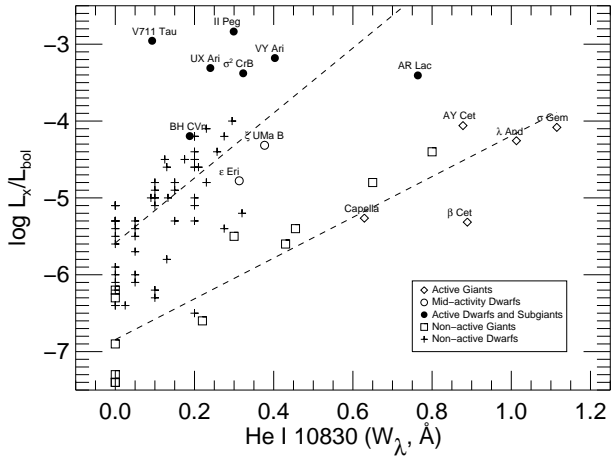}
   \caption{$L_{\rm X}/L_{\rm bol}$
     vs the equivalent width of  the \ion{He}{i} 10830~\AA\
   triplet. Non-active stars are included from \citet{zar86}. Dashed
     lines indicate a linear fit to all giants ($r=0.92$),
     and a linear fit to dwarfs and subgiants ($r=0.65$).}  
   \label{fig:helx}
\end{figure}

If we restrict the analysis to giants only, there is a clear
relation: an increasing EUV or X-ray luminosity
(Figs.~\ref{fig:heeuv}, \ref{fig:helx}) corresponds to
increasing equivalent widths of the \ion{He}{i} 10830~\AA\
line. Such a relation was found earlier by \citet{zir82} and
\citet{smi83} using a large sample of giants observed 
with mostly photographic plates and X-ray fluxes from the 
Einstein satellite.
\citet{smi83} included several RS~CVn stars
in his tabulation, using EUV measurements from \citet{obr80}.  Smith  reported
a relationship for giant stars, as did \citet{obr86}.
Our sample of active giants contains 
observations that are generally of better quality in both the 10830~\AA\ line and
the X-ray flux. Moreover we measure the  EUV flux directly, which
is close to the helium edge ($\sim$504~\AA); therefore it should
be better related to the 10830~\AA\ flux if controlled by
photoionization rather than radiation from
shorter wavelengths. 
The flux of the \ion{He}{ii}~304~\AA\ line could be added to the flux
of the ionizing
EUV radiation, but the relations do not change significantly since the  304~\AA\ flux is
generally much less than the total EUV flux. One of the stars (AY Cet) has
no direct measurement of the 304~\AA\ line. 
The quality of the spectra
at longer wavelengths allows us to approximate the flux
in the whole band 170--504~\AA\ for only a few cases. For Capella,
the most extreme case, the 304~\AA\ line flux represents only
$\sim$20\% of the total flux (in photon units) in the 80--504~\AA\ band.
Therefore we consider the \ion{He}{ii}~304~\AA\ line
to be only a minor contributor to the excitation of the 10830~\AA\
line in these stars.

We have also compared the EUV flux to that of the
\ion{He}{ii}~304~\AA\ line. If a PR mechanism dominates the 304~\AA\
formation, 
we might find higher fluxes in the 304~\AA\ line for higher EUV
luminosities. However,  collisional processes could produce a similar
dependence. The strength of the \ion{He}{ii}~304~\AA\ line is
clearly related to the EUV flux, but it is also strongly correlated
with the \ion{C}{iv} flux.  This is not 
surprising since collisions dominate the production of both the EUV
and \ion{C}{iv} emission.  It is not possible with these spectra to evaluate
the contribution of the PR mechanism to the formation of the 304~\AA\ line.
Since the
\ion{He}{ii}~1640~\AA\ line shares a level with the 304~\AA\ line, we
might use it to test the formation of the latter. However such an exercise
was carried out for Capella \citep{dup93}, and the 304~\AA\ line had
only $\sim 30$\% of the expected flux for the measured
1640~\AA\ line, perhaps resulting from  interstellar medium absorption
or opacity effects in the star itself.

The two competing mechanisms for formation of the 10830~\AA\ line 
could both be  important in the
stars of the sample. Since they have  high temperatures in the
transition regions, collisional population of the levels
involved in the line formation can occur. Stars in the sample also have 
large X-ray and EUV fluxes causing photoionization of  \ion{He}{i}
which is followed by recombination, populating the same levels.
However the model proposed for dwarf stars by \citet{and95}, with
collisional effects dominating 
for the active stars, 
predicts the maximum equivalent widths observed here.  Their model
is also consistent with the observed filled-in emission with increasing EUV flux.

\begin{figure}
   \centering
   \hspace{10mm}
   \includegraphics[width=0.5\textwidth]{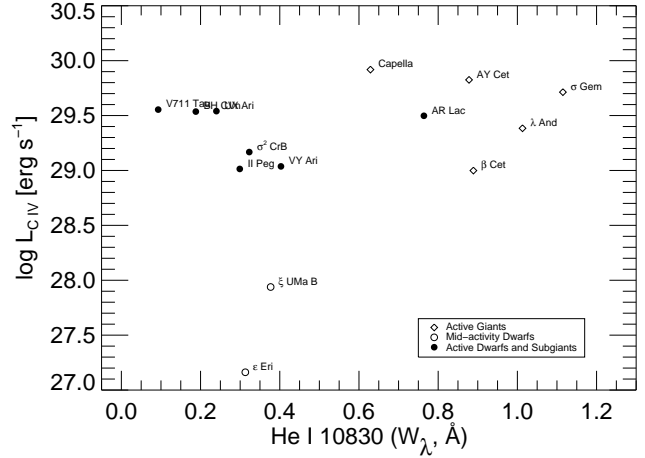}
   \caption{\ion{C}{iv} ($\lambda$1550) luminosity vs \ion{He}{i}
     $\lambda$10830 equivalent width.  Fluxes for \ion{C}{iv} are taken
     from {\it IUE} spectra \citep[see][]{sanz02,sanz03}.}
   \label{fig:c4he}
\end{figure}

\begin{figure}[t]
   \centering
   \hspace{10mm}
   \includegraphics[width=0.5\textwidth]{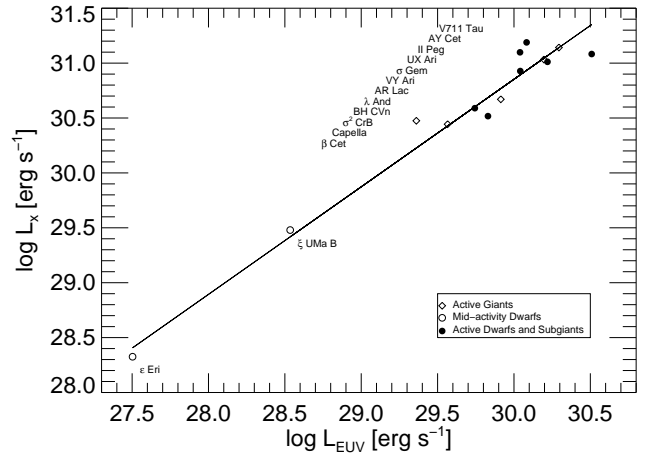}
   \caption{X-ray vs EUV luminosity for the stars in this sample.  Not
   surprisingly, a correlation is present ($r=0.98$), since both
   spectral regions 
are dominated by collisionally excited line emission formed at similar
high temperatures.}
   \label{fig:euvx}
\end{figure}

\begin{figure}
   \centering
   \hspace{10mm}
   \includegraphics[width=0.5\textwidth]{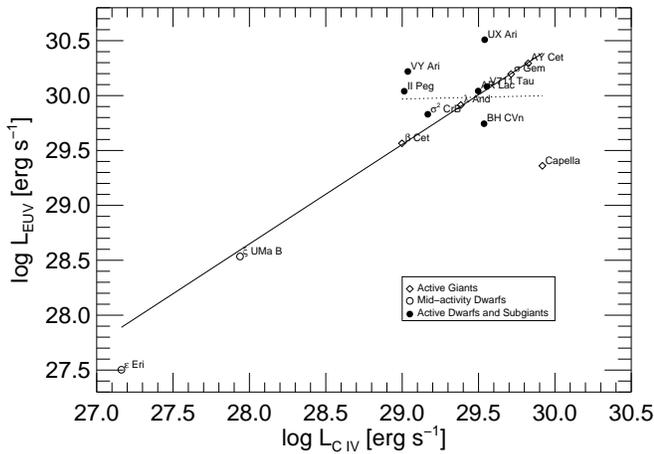}
   \caption{EUV luminosity vs. \ion{C}{iv}
     ($\lambda$1550) luminosity. A fit to all the stars is shown in
     solid line ($r=0.86$), while the dashed line excludes 
     $\epsilon$ Eri and $\xi$ UMa B from the fit ($r=0.03$).} 
   \label{fig:c4euv}
\end{figure}

The wings of the \ion{He}{i}
10830~\AA\ line in giants are broader than in dwarf
stars which is indicative of mass
motions and  outflows that are likely present in the upper 
chromosphere or at the base of coronal loops. 
\citet{kat98} proposed that the formation of EUV lines and the \ion{He}{i}
10830~\AA\ line in stars such as Capella results from a
magnetized stellar wind forming a shock in the corona of the companion
star. Now we know that 
the EUV fluxes found in Capella are similar to those found in other
single active stars \citep{sanz03}, and therefore 
strong EUV emission is not linked to
binarity.  The most logical explanation is that EUV lines are formed
in dense coronal loops hotter than those found in the Sun, and the
10830~\AA\ line is formed as a consequence of the EUV flux, and 
additionally through collisional effects in the hot environment of the
transition region. 

An uncertainty remains whether the \ion{He}{i} absorption for the 
giants is controlled by the EUV/X-ray flux, or whether it 
simply reflects a general increase in  magnetic activity represented
by dense regions 
and subsequent radiative losses from  the star.  \citet{obr86} 
suggested that the
Helium I line strength in giants and supergiants correlated as well
with other lines 
known to be collisionally excited, and hence is, in fact, independent of the
EUV/X-ray flux. Our small sample shows no such correlation
(Fig.~\ref{fig:c4he}). For the 
dwarf stars, the luminosity
of the collisionally excited \ion{C}{iv} multiplet
($\lambda$1550) shows a saturation 
in \ion{C}{iv} and a limiting value (400 m\AA) of the \ion{He}{i} equivalent
width (Fig.~\ref{fig:heeuv} and Fig.~\ref{fig:c4he}).  The giant stars
show no correlation ($r=-0.31$)
between \ion{C}{iv} and the \ion{He}{i} equivalent width
(Fig.~\ref{fig:c4he}), suggesting that collisions are not significant
in increasing the strength of \ion{He}{i}. The radiation field must
therefore influence the line strength of \ion{He}{i}.  
While the X-ray and EUV fluxes themselves are correlated
(Fig.~\ref{fig:euvx}), the 
EUV and \ion{C}{iv} fluxes are not correlated for the active dwarfs
and giants (Fig.~\ref{fig:c4euv}).  
The small scale size of the high temperature EUV-emitting regions
\citep{dup93,sanz02,sanz03} suggests they remain
of comparable small size in  stellar chromospheres of varying
dimensions.

\section{Conclusions}
The \ion{He}{i} $\lambda$10830 line responds differently
to the EUV radiation field between the
dwarf  and giant stars in our sample. Active dwarf 
stars reach a 'saturated' equivalent width in $\lambda$10830 in 
the presence of  
a strong radiation field.  This behavior appears 
consistent with model calculations in which high chromospheric
densities allow collisional excitation to dominate
photoionization/recombination processes in forming the line.  
Giant stars, with lower chromospheric densities than dwarfs, show
increased \ion{He}{i} absorption related to an increased EUV
radiation field and the absorption is  strengthened by an extended
expanding atmosphere scattering the line. The \ion{He}{i} line in
giant stars has a photoionization-recombination component
that appears to dominate the line-forming process. Detailed radiative
transfer calculations 
would be helpful to assess the contribution of collisions to
line formation in the giant stars.

\begin{acknowledgements}
We are grateful to Ilya Ilyin for assistance with the observing 
activities at the NOT.  In addition, we thank the Spanish TAC for the  
allocation of telescope time at the NOT. 
We acknowledge the anonymous referee, who helped greatly in clarifying the
analysis and presentation of the results.
This work made use of the Smithsonian/NASA Astrophysics Data System and SIMBAD, a
product of the Centre de Donn\'ees astronomique de Strasbourg.
\end{acknowledgements}

\bibliographystyle{aa}


\end{document}